\pdfoutput=1
\documentclass{optica-article}

\journal{opticajournal}

\articletype{Research Article}

\usepackage{amsmath}
\usepackage{siunitx}
\usepackage{float}

\begin{document}

\title{Broadband High-Level Squeezed Light using Waveguide Optical Parametric Amplifiers with External Dispersion Compensation }

\author{Takumi Suzuki,\authormark{1,$\dag$} Shotaro Oki,\authormark{1,$\dag$} Kazuki Hirota,\authormark{1} Takaya Hoshi,\authormark{1} Ryuhoh Ide,\authormark{1} Takahiro Kashiwazaki,\authormark{2} Taichi Yamashima,\authormark{2} Asuka Inoue,\authormark{2} Takeshi Umeki,\authormark{2} Mamoru Endo,\authormark{1,3} and Akira Furusawa\authormark{1,3,4,*}}

\address{\authormark{1}Department of Applied Physics, School of Engineering, The University of Tokyo, 7-3-1 Hongo, Bunkyo-ku, Tokyo 113-8656, Japan\\
\authormark{2}Device Technology Labs, NTT, Inc., 3-1 Morinosato Wakamiya, Atsugi, Kanagawa 243-0198, Japan\\
\authormark{3}RIKEN Center for Quantum Computing, 2-1 Hirosawa, Wako, Saitama 351-0198, Japan\\
\authormark{4}OptQC Corp., 3-28-13 Nishi-Ikebukuro, Toshima-ku, Tokyo 171-0021, Japan\\
\authormark{$\dag$}These authors contributed equally to this work.}

\email{\authormark{*}akiraf@ap.t.u-tokyo.ac.jp}

\begin{abstract*}
We demonstrate broadband phase-sensitive amplification (PSA) measurement of squeezed light generated by a waveguide optical parametric amplifier (OPA) with external dispersion compensation. In broadband systems, group velocity dispersion (GVD) induces a frequency-dependent rotation of the squeezing axis, which limits the observable bandwidth in PSA measurements. While dispersion compensation between two OPAs has been demonstrated previously, we show that the quantity to be compensated is not the material group-delay dispersion $D$ itself but a gain-dependent effective value $D\tanh r/(2r)$, where $r$ is the squeezing parameter, and realize the compensation with free-space fused-silica plates whose insertion loss is $0.4\%$. As a result, we observe a maximum squeezing of $5.9\pm0.2$~dB near the carrier frequency and more than 5~dB of squeezing up to a frequency offset of $5.9\pm0.1$ THz from the carrier. Furthermore, squeezing below the shot-noise level is confirmed up to a frequency offset of $6.9\pm0.1$ THz from the carrier, corresponding to the accessible phase-matching bandwidth of the waveguide OPA. Our results establish a practical method for broadband characterization of squeezed light and provide a key step toward ultrafast continuous-variable quantum information processing.
\end{abstract*}

\section{Introduction}

Squeezed light is a nonclassical optical field that exhibits reduced quantum noise below the standard quantum limit and plays a fundamental role in continuous-variable quantum information processing and precision measurements \cite{Braunstein2005, Giovannetti2004, Andersen2016}. 
In particular, squeezed states serve as essential resources for quantum teleportation \cite{Furusawa1998, Braunstein1998}, cluster-state generation \cite{Asavanant2019, Larsen2019}, and measurement-based quantum computation \cite{Menicucci2006, Raussendorf2001,Raussendorf2003}. 
Therefore, the realization of high-level squeezing over a broad spectral bandwidth is a crucial requirement for advancing continuous-variable quantum technologies.

Traditionally, high levels of squeezing have been achieved using optical parametric oscillators (OPOs) \cite{Vahlbruch2016}. However, their cavity structure inherently limits the bandwidth to the order of GHz. In contrast, waveguide optical parametric amplifiers (OPAs) do not rely on optical cavities and thus can, in principle, generate squeezed light over the entire phase-matching bandwidth, which can extend to the THz regime \cite{Kashiwazaki2021,Nehra2022}. Recent experiments using periodically poled lithium niobate (PPLN) waveguides have demonstrated broadband squeezing spanning 2.5 THz with a squeezing level of 6 dB \cite{Kashiwazaki2020}.

To characterize such broadband squeezed light, phase-sensitive amplification (PSA) has been proposed as a powerful measurement technique \cite{Shaked2018, Kalash2023}. In PSA, a specific quadrature component of the optical field is selectively amplified by an OPA, and its spectral noise can be directly measured using an optical spectrum analyzer. However, in OPAs, group-velocity dispersion (GVD) causes a frequency-dependent rotation of the squeezing axis, leading to a mismatch between the squeezing direction and the measurement axis \cite{Kashiwazaki2021}. As a result, the observed squeezing is degraded at large frequency offsets, significantly limiting the effective measurement bandwidth.

Previous approaches have attempted to compensate for this effect by introducing dispersion-compensating fibers between OPAs \cite{Takanashi2020}. However, such methods do not fully cancel the dispersion accumulated inside the waveguides and often introduce additional optical loss. Consequently, broadband PSA measurements covering the entire phase-matching bandwidth have not yet been achieved.

In this work, we demonstrate broadband PSA measurement of squeezed light by combining a low-loss free-space configuration with external dispersion compensation. While this two-OPA architecture follows Ref.~\cite{Takanashi2020}, we develop a theoretical model that describes the combined effect of squeezing and dispersion in a unified framework and derive the optimal dispersion compensation condition, which shows that the quantity to be compensated is a gain-dependent effective dispersion $D\tanh r/(2r)$ rather than the material dispersion itself. Experimentally, we employ a two-stage waveguide OPA system with fused silica-based dispersion compensation inserted between the stages. We observe up to $5.9\pm0.2$~dB of squeezing near the carrier frequency and more than 5~dB of squeezing up to a frequency offset of $5.9\pm0.1$ THz from the carrier. Moreover, squeezing below the shot-noise level is confirmed up to a frequency offset of $6.9\pm0.1$ THz from the carrier, corresponding to the accessible phase-matching bandwidth. These results provide a practical method for broadband characterization of squeezed light and open a pathway toward ultrafast all-optical quantum information processing.

\section{Theory}
\begin{figure}[htbp]
  \centering
  \includegraphics[width=13cm]{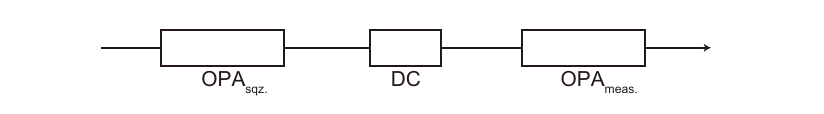}
\caption{Conceptual model of the two-stage OPA scheme. A squeezed vacuum generated in OPA$_{\mathrm{sqz}}$ propagates through a dispersion-compensation element (DC) and is measured by phase-sensitive amplification in OPA$_{\mathrm{meas}}$. The DC element cancels the frequency-dependent quadrature rotation accumulated between the two OPAs.}\label{fig.concept}
\end{figure}

We consider a two-stage optical parametric amplification system for squeezed vacuum generation and all-optical measurement, as illustrated in Fig.~\ref{fig.concept}. 
The first OPA (OPA$_\mathrm{sqz.}$) operates in a relatively low-gain regime and is used to generate a squeezed vacuum state. 
The second OPA (OPA$_\mathrm{meas.}$) operates in a higher-gain regime and is used for phase-sensitive amplification (PSA) measurement of the generated state. 
The two OPAs are connected via a dispersive optical path, and an additional dispersion compensation element (DC) is inserted between them. 
The measurable PSA bandwidth is determined by the interplay between the squeezing operations and the frequency-dependent quadrature rotation induced by dispersion.

We first describe the principle of squeezed light generation and PSA measurement using OPAs. 
We then analyze the effect of dispersion as a frequency-dependent quadrature rotation. 
Finally, we derive the condition for optimal dispersion compensation in a two-stage OPA system, which maximizes the measurable PSA bandwidth.

\subsection{Principle of squeezed light generation and PSA measurement with OPAs}

In this work, two OPAs are employed with distinct roles. The first OPA operates at a relatively low gain and is used to generate a squeezed vacuum state. The second OPA operates in a higher-gain regime and is used for PSA measurement.

We represent the quadrature vector as
\begin{equation}
v = \begin{pmatrix} x \\ p \end{pmatrix}.
\end{equation}

An ideal degenerate OPA performs the squeezing transformation
\begin{equation}
v \rightarrow S(r)v, \quad
S(r)=
\begin{pmatrix}
e^{r} & 0 \\
0 & e^{-r}
\end{pmatrix},
\end{equation}
where $r$ is the squeezing parameter.
In particular, when the input is a vacuum state, the output state becomes a squeezed vacuum state, in which the quantum noise is reduced in one quadrature and enhanced in the orthogonal one. This configuration is used for the OPA dedicated to squeezed vacuum generation (hereafter denoted as OPA$_{\mathrm{sqz}}$).

In contrast, the OPA can also be used for measurement (hereafter denoted as OPA$_{\mathrm{meas}}$). In the high-gain condition ($e^{2r} \gg 1$), the above transformation strongly amplifies one quadrature while suppressing the orthogonal one.
By performing homodyne detection with a local oscillator phase-matched to the amplified quadrature, the quadrature amplitude along the amplification axis can be directly measured \cite{Caves1982}. In fact, quadrature measurements with a bandwidth of approximately the tens of GHz have been demonstrated \cite{Inoue2023,Kawasaki2024,Kawasaki2025}.
However, it is generally difficult to extend homodyne detection to broader bandwidths due to the limitations of electronic circuits. Therefore, for measurements in the THz regime and beyond, optical spectrum measurements based on direct power detection are commonly employed \cite{Kashiwazaki2021,Nehra2022}.
For simplicity, the dispersion inside OPA$_{\mathrm{meas}}$ is neglected in this introductory subsection; it is fully incorporated in Sec.~2.4.

In this case, the amplified quadrature dominates the output intensity. The intensity, corresponding to the photon-number-like quantity, is given by
\begin{align}
\langle N_{\mathrm{out}}\rangle =& e^{2r}\langle\hat{x}_{\mathrm{in}}^2\rangle + e^{-2r}\langle\hat{p}_{\mathrm{in}}^2\rangle \\
\approx& e^{2r}\langle\hat{x}_{\mathrm{in}}^2\rangle
\end{align}
in the high-gain condition, where $\langle\hat{x}_{\mathrm{in}}^2\rangle$ denotes the noise variance (power spectral density) of the selected quadrature.
Therefore, the selected quadrature can be effectively read out via direct power (optical spectrum) measurement without the need for an explicit local oscillator.

\subsection{Dispersion-induced quadrature rotation}
The phase accumulated in a dispersive medium can be expanded in powers of the frequency offset. 
Odd-order terms correspond to a global phase shift and group delay, which do not affect the quadrature basis. 
In contrast, even-order terms lead to a frequency-dependent rotation in phase space. 
In this work, we retain the second-order term (group delay dispersion, GVD), which provides the dominant contribution to the quadrature rotation.

In a dispersive medium, the wave number is expanded as
\begin{equation}
k(\omega_0+\Omega) = k_0 + k_1\Omega + \frac{1}{2}k_2\Omega^2 + \cdots,
\end{equation}
where $k_2$ corresponds to the GVD.
After propagation over a length $L$, the accumulated phase is
\begin{equation}
\phi(\Omega) = \frac{1}{2}D\,\Omega^2,
\end{equation}
where $D = k_2 L$ is the group delay dispersion (GDD).

This phase shift corresponds to a rotation in quadrature space,
\begin{equation}
v_{\mathrm{out}} = R(\phi)v_{\mathrm{in}},
\end{equation}
with
\begin{equation}
R(\phi)=
\begin{pmatrix}
\cos\phi & -\sin\phi \\
\sin\phi & \cos\phi
\end{pmatrix}.
\end{equation}

Since $\phi(\Omega)$ depends on frequency, the squeezing axis rotates across the spectrum, which degrades the observable squeezing in PSA measurements.

\subsection{Combined effect of squeezing and dispersion}

In an OPA, squeezing and dispersion occur simultaneously.
The combined transformation can be written as
\begin{equation}
A(r,\theta) = \exp
\begin{pmatrix}
r & -\theta \\
\theta & -r
\end{pmatrix},
\end{equation}
where
\begin{equation}
\theta(\Omega) = \frac{1}{2}D\,\Omega^2.
\end{equation}
Using the Bloch--Messiah decomposition \cite{Braunstein2005-2}, this transformation can be expressed as
\begin{equation}
A(r,\theta) = R(\alpha)S(a)R(\beta),
\end{equation}
where $a$ is the effective squeezing parameter and $\alpha(\Omega), \beta(\Omega)$ describe frequency-dependent rotations. 
The relation $\alpha=\beta$ follows from the symmetry of the generator $B$ (a detailed derivation is given in Appendix~\ref{app:bloch-messiah}, where the same quantities are denoted by $r'$ and $\phi$, respectively).

These parameters can be obtained analytically as
\begin{align}
a &= \mathrm{arcsinh}\left(\frac{r}{k}\sinh k\right), \\
\alpha = \beta &= \frac{1}{2}\arctan\left(\frac{\theta }{k}\tanh k\right),
\end{align}
where
\begin{equation}
k = \sqrt{r^2 - \theta^2}.
\end{equation}
The above expressions assume $r > \theta$, such that $k = \sqrt{r^2 - \theta^2}$ remains real. 
This condition is satisfied in the vicinity of the carrier frequency, where the present analysis is focused. For the parameters used in this work, the condition $r = \theta$ is reached at frequency offsets on the order of several THz (approximately 5--6~THz), beyond which the present analytical solution changes (see Appendix A.2).

\begin{figure}[htbp]
  \centering
  \includegraphics[width=13cm]{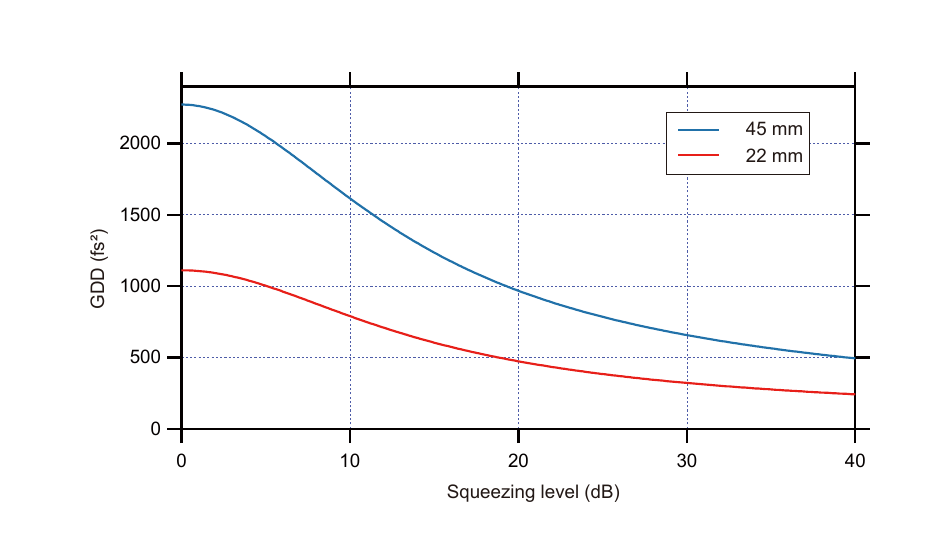}
\caption{Effective GDD of the squeezed quadrature versus squeezing level $10\log_{10}e^{2r}$ for the two OPAs ($22~\mathrm{mm}$, $D_{\mathrm{sqz.}}=2222~\mathrm{fs^2}$; $45~\mathrm{mm}$, $D_{\mathrm{meas.}}=4545~\mathrm{fs^2}$). The effective GDD scales as $\tfrac{D}{2}\,\tanh r/r$ [Eq.~\eqref{eq:alpha-low-freq}] and decreases with increasing gain.}\label{fig.simulation_22mm_45mm}
\end{figure}

The second derivative of $\alpha$, i.e., the effective GDD as a function of the squeezing level ($10\log_{10} e^{2r}$~[dB]), is plotted in Fig.~\ref{fig.simulation_22mm_45mm}. 
Here, we consider two PPLN waveguide OPAs used in the experiment discussed later, with the lengths and total GDD values of (22 mm, 2222~fs$^2$) and (45 mm, 4545~fs$^2$), respectively \cite{Zelmon1997}. 
As can be seen from the figure, the effective GDD strongly depends on the squeezing level of the OPA.

In particular, expanding $\alpha(\Omega)$ around the carrier frequency yields
\begin{equation}
\alpha(\Omega) \approx \frac{D}{4}\frac{\tanh r}{r}\Omega^2.
\label{eq:alpha-low-freq}
\end{equation}
This frequency-dependent rotation is the primary mechanism limiting the OPA bandwidth.

\subsection{Dispersion compensation}

For a two-stage OPA system with external DC, the total transformation is
\begin{equation}
v_{\mathrm{out}} =
A(r_\mathrm{meas.},\theta_\mathrm{meas.})
R(\phi_{\mathrm{dc}})
A(r_\mathrm{sqz.},\theta_\mathrm{sqz.})
v_{\mathrm{in}}.
\label{eq:two-stage-total}
\end{equation}

Applying the Bloch--Messiah decomposition to each stage and combining the three consecutive rotations in the middle into the single angle $\alpha_{\mathrm{tot}}\equiv\alpha_\mathrm{meas.}+\phi_{\mathrm{dc}}+\alpha_\mathrm{sqz.}$, we obtain
\begin{equation}
v_{\mathrm{out}} =
R(\alpha_\mathrm{meas.})\,S(a_\mathrm{meas.})\,
R(\alpha_{\mathrm{tot}})\,
S(a_\mathrm{sqz.})\,R(\alpha_\mathrm{sqz.})
v_{\mathrm{in}}.
\label{eq:two-stage-bm}
\end{equation}
The input rotation $R(\alpha_\mathrm{sqz.})$ defines the reference phase of the squeezed quadrature and corresponds to a choice of basis. 
The output rotation $R(\alpha_\mathrm{meas.})$ does not affect the experimental results in this work, since the measurement is based on optical power and is insensitive to the quadrature phase. Even in homodyne detection, it can be absorbed into the phase of the local oscillator.
Therefore, both input $R(\alpha_\mathrm{sqz.})$ and output $R(\alpha_\mathrm{meas.})$ can be omitted without loss of generality.

In contrast, the total frequency-dependent rotation inside the system,
\begin{equation}
\alpha_{\mathrm{tot}}(\Omega)
=
\alpha_\mathrm{sqz.}(\Omega)
+
\phi_{\mathrm{dc}}(\Omega)
+
\alpha_\mathrm{meas.}(\Omega),
\label{eq:alpha-tot}
\end{equation}
cannot be removed and determines the mismatch between the amplified quadrature and the measurement basis.

The measured squeeze bandwidth is maximized when
\begin{equation}
\alpha_{\mathrm{tot}}(\Omega) \approx 0.
\label{eq:opt-condition}
\end{equation}

Although the optimal $D_\mathrm{dc}$ can be calculated numerically, using the low-frequency approximation, the optimal dispersion compensation is given by
\begin{equation}
D_{\mathrm{dc}}
\approx
- D_\mathrm{sqz.} \frac{\tanh r_\mathrm{sqz.}}{2 r_\mathrm{sqz.}}
- D_\mathrm{meas.} \frac{\tanh r_\mathrm{meas.}}{2 r_\mathrm{meas.}}.
\label{eq:opt-Ddc}
\end{equation}

For example, using the nominal parameters of the experiment discussed later in this paper ($r_\mathrm{sqz.} = 1.497$ (13 dB), $r_\mathrm{meas.} = 2.993$ (26 dB), and $D_\mathrm{sqz.} = 2222~\mathrm{fs^2}, D_\mathrm{meas.} = 4545~\mathrm{fs^2}$), the optimal compensation is estimated to be $D_{\mathrm{dc}} = -1427~\mathrm{fs^2}$. 
This analytical value coincides with the result obtained by numerically minimizing $\alpha_{\mathrm{tot}}(\Omega)$ near the carrier frequency ($-1427~\mathrm{fs^2}$), confirming that the low-frequency approximation reproduces the exact quadrature-rotation curvature at zero frequency.

Alternatively, the dispersion compensation can be optimized numerically by minimizing the residual rotation over a finite bandwidth. 
For instance, under the same conditions, minimizing the integrated squared rotation residual up to a frequency offset of $\pm 5$~THz yields $D_{\mathrm{dc}} = -1485~\mathrm{fs^2}$. 
Depending on the application, the appropriate compensation value can be selected. 
In addition, fine adjustment of the squeezing level provides another degree of freedom for optimizing the overall performance.

\subsection{Simulation of PSA spectrum with dispersion compensation}

\begin{figure}[htbp]
  \centering
  \includegraphics[width=13cm]{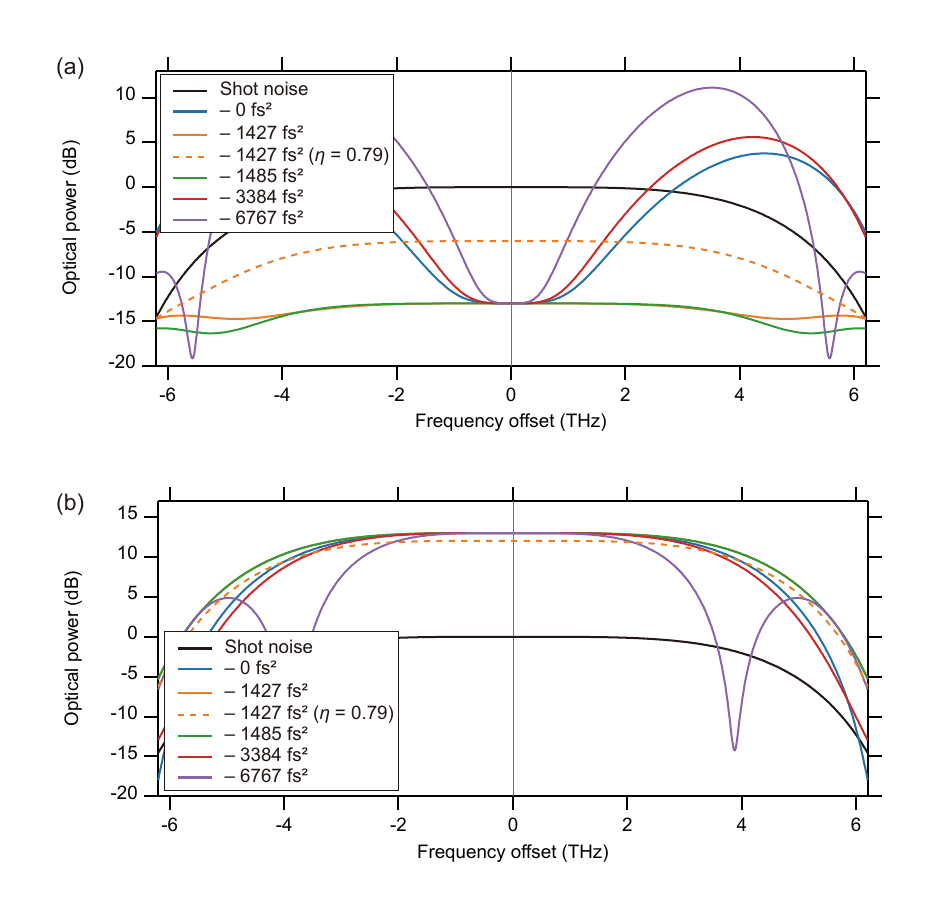}
\caption{Simulated PSA spectra for several dispersion-compensation values $D_{\mathrm{dc}}$: $0$, $-1427$ (analytical optimum), $-1485$ ($\pm5~\mathrm{THz}$ optimum), $-3384$ (zero-gain limit), and $-6767~\mathrm{fs^2}$ (full material compensation). (a) Squeezed and (b) anti-squeezed quadratures. Solid black: shot noise, i.e.\ the vacuum amplified by OPA$_\mathrm{meas.}$, to whose value at the carrier all spectra are normalized. The dashed curve is the analytical optimum recomputed with the measured detection efficiency $\eta=0.79$. Proper compensation broadens the squeezed band in (a); (b) is far less sensitive to dispersion and to loss.}\label{fig.simulation_result}
\end{figure}

To quantitatively evaluate the effect of dispersion compensation, we numerically simulate the measurement performed with an optical spectrum analyzer (OSA). We first assume an ideal lossless system in order to isolate the intrinsic effect of dispersion-induced quadrature rotation, and then repeat the calculation with the total detection efficiency measured in this work.

The squeezing parameters and the dispersions of the two OPAs are fixed to the experimental values, while the amount of dispersion compensation ($D_\mathrm{dc}$) is varied. Specifically, we consider five representative cases: no compensation ($D_{\mathrm{dc}} = 0~\mathrm{fs}^2$), optimal compensation based on the analytical estimate ($D_{\mathrm{dc}} = -1427~\mathrm{fs}^2, -1485~\mathrm{fs}^2$), the optimal compensation in the zero-gain limit ($D_{\mathrm{dc}} = -3384~\mathrm{fs}^2$, which corresponds to $-(D_\mathrm{sqz.}+D_\mathrm{meas.})/2$ obtained by taking $\tanh r/r\to 1$ in Eq.~\eqref{eq:opt-Ddc}), and full compensation of the material dispersion of both OPAs ($D_{\mathrm{dc}} = -(D_\mathrm{sqz.} + D_\mathrm{meas.}) = -6767~\mathrm{fs}^2$). 
By comparing these cases, we clarify how the choice of dispersion compensation affects the PSA gain spectrum and bandwidth.

Figure~\ref{fig.simulation_result} (a) shows the optical spectra obtained when the squeezed quadrature is amplified in the second OPA. From these plots, it can be seen that, in the absence of dispersion compensation, as well as in the cases where the dispersion of the two OPAs is simply compensated or where the optimal value in the zero-gain limit (i.e., half of the total dispersion of the two OPAs) is used, a high level of squeezing is obtained only in the vicinity of the carrier frequency.
In contrast, when the value calculated using the analytical approximation is applied, a significant improvement in the squeezing bandwidth is observed. Furthermore, by tuning the compensation to minimize the residual quadrature rotation over a wider frequency range, the squeezing level at higher offset frequencies, around 5~THz, is further improved.

These results indicate that, for practical purposes, the analytically estimated value from Eq.~\eqref{eq:opt-Ddc} already provides sufficient performance. However, additional fine-tuning enables further enhancement of the squeezing properties in the high-frequency region. Such fine-tuning can be implemented by adjusting the dispersion compensation element, but it can also be achieved by slightly modifying the squeezing parameters. From a practical standpoint, the latter approach is advantageous.

In addition, Fig.~\ref{fig.simulation_result} (b) corresponds to the case where the amplification axis of the second OPA is rotated by $90^\circ$, i.e., the anti-squeezed quadrature is measured. In this configuration, the effect of dispersion is less pronounced than in the squeezed quadrature. This asymmetry arises because the squeezed quadrature is exponentially sensitive to phase mismatch, whereas the anti-squeezed quadrature is comparatively robust.

Nevertheless, it can be seen that appropriate dispersion compensation still broadens the flat spectral region even in this case.

\paragraph{Effect of optical loss.}
The dashed orange curves in Fig.~\ref{fig.simulation_result} show the same calculation at the analytical optimum $D_{\mathrm{dc}}=-1427~\mathrm{fs^2}$, but including the total detection efficiency $\eta=0.79$ measured for the present setup (Sec.~3.1, Table~\ref{tab:loss-budget}). Loss is modeled as a beam splitter of transmittance $\eta$ inserted after OPA$_\mathrm{sqz.}$, i.e.\ $M\to\eta M+(1-\eta)I$ for the covariance matrix. At the carrier frequency the squeezed quadrature rises from $-13.0$~dB to $-6.0$~dB, while the anti-squeezed quadrature falls only from $+13.0$~dB to $+12.0$~dB. This strong asymmetry---seven decibels against one---is the reason why the lossless simulation and the measured squeezing level differ so much: the squeezed quadrature is exponentially sensitive to admixed vacuum, whereas the anti-squeezed quadrature is not. The predicted level of $-6.0$~dB agrees with the measured value reported in Sec.~3.2 within the experimental uncertainty.

Importantly, the dashed curve remains flat over essentially the same bandwidth as the lossless curve at the same $D_{\mathrm{dc}}$: optical loss lowers the achievable squeezing level but leaves the optimal amount of compensation almost unchanged. Quantitatively, the optimal dispersion compensation is reduced compared to the lossless case, because vacuum fluctuations introduced along the propagation experience only a fraction of the total dispersion, leading to an effectively smaller accumulated phase rotation; for the parameters of this experiment the shift amounts to only a few femtoseconds squared (Sec.~3.2). The details can be found in Appendix~\ref{app:lossy}.

\section{Experiment}
\subsection{Experimental apparatus}
\begin{figure}[htbp]
  \centering
  \includegraphics[width=13cm]{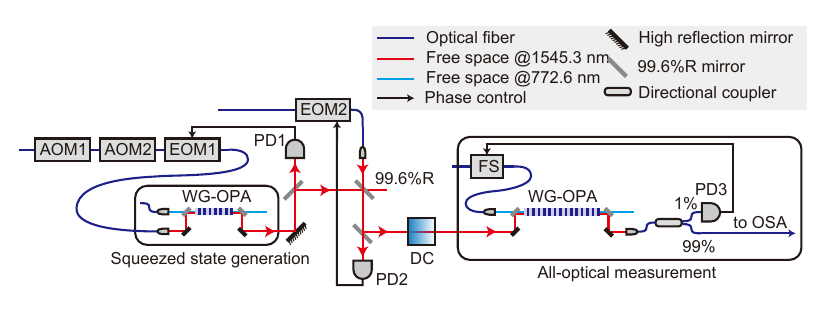}
\caption{Experimental setup. Squeezed light from OPA$_{\mathrm{sqz}}$ ($22~\mathrm{mm}$) is measured by PSA in OPA$_{\mathrm{meas}}$ ($45~\mathrm{mm}$), both pumped at $388~\mathrm{THz}$ (signal $194~\mathrm{THz}$). Fused-silica plates at the Brewster angle provide dispersion compensation. A probe beam (locked via a $99.6\%$ mirror) and a separate lock beam (stabilized through an electro-optic modulator (EOM)) reference the two OPAs. Output is recorded with an optical spectrum analyzer (OSA).}\label{fig.setup}
\end{figure}
The experimental setup is shown in Fig.~\ref{fig.setup}. Both OPAs are implemented using PPLN waveguides, with lengths of 22~mm and 45~mm for squeezed light generation (OPA$_{\mathrm{sqz}}$) and measurement (OPA$_{\mathrm{meas}}$), respectively.  Further details of the waveguides can be found in \cite{Kashiwazaki2020}. Both waveguides are pumped by continuous-wave light at 388~THz. The pump powers are adjusted so that the parametric gains are close to the nominal values $r_\mathrm{sqz.}=1.497$ ($13~\mathrm{dB}$) and $r_\mathrm{meas.}=2.993$ ($26~\mathrm{dB}$) used in the numerical simulations of the previous section; the values realized for the data of Fig.~\ref{fig:experimental_result}(b) are $13.2\pm0.4~\mathrm{dB}$ and $25.3\pm0.4~\mathrm{dB}$. The quoted uncertainty is the scatter of the measured gains about the expected scaling with the square root of the pump power, evaluated over the nine data sets recorded in this work. 

Since parametric amplification is phase-sensitive and the squeezed vacuum state itself does not provide a phase reference, a classical probe beam (Probe) is co-propagated with the signal field. The phase information is obtained from classical parametric amplification between the probe beam and the pump, and is used for feedback control. Specifically, the phase relationship between the pump and the probe (and hence between the squeezed field and the probe) is extracted from the modulation signal of the probe beam transmitted through a 99.6\% mirror placed between the two OPAs. This signal is fed back to a phase modulator in the probe path. For the second OPA, an independent locking scheme is required. 
If the amplified probe beam from the first OPA is directly used as a phase reference for the second OPA, the squeezed vacuum co-propagating with the probe is also strongly amplified along the same quadrature, contaminating the modulation sideband and degrading the signal-to-noise ratio of the error signal.
To overcome this issue, an additional classical beam (lock beam) is introduced, as shown in Fig.~\ref{fig.setup}. The relative phase between the probe and the lock beam is stabilized using an electro-optic modulator (EOM), and the phase of the second OPA is locked via the phase relationship between the lock beam and the pump.

To compensate for dispersion, fused silica plates are inserted between the two OPAs at the Brewster angle. The group velocity dispersion of fused silica at 194~THz is approximately $-27.36~\mathrm{fs^2/mm}$ \cite{Malitson1965}, with a refractive index of $n=1.444$, so that the Brewster angle is $55.3^{\circ}$, the internal propagation angle is $34.7^{\circ}$, and the optical path length is $1.216$ times the plate thickness. 
Hereafter we report $D_{\mathrm{DC}}\equiv|D_{\mathrm{dc}}|$ to denote the magnitude of the negative dispersion introduced by the fused-silica plates. With this convention, the analytical optimum of Eq.~\eqref{eq:opt-Ddc}, evaluated with the parametric gains realized in the experiment, corresponds to $D_{\mathrm{DC}}  = 1440\pm20~\mathrm{fs^2}$.

Since fused-silica plates are available only in discrete thicknesses, the accessible values of $D_{\mathrm{DC}}$ are discrete as well; with the $20$- and $30~\mathrm{mm}$-thick plates available to us, $D_{\mathrm{DC}}=1331\pm8~\mathrm{fs^2}$, obtained with a pair of $20~\mathrm{mm}$ plates and limited by the $1^{\circ}$ accuracy of the angular setting, is the accessible value closest to the optimum; the value $1997\pm12~\mathrm{fs^2}$ used in Fig.~\ref{fig:experimental_result}(c) corresponds to a pair of $30~\mathrm{mm}$ plates. The two plates of a pair are chosen to have equal thickness and are mounted anti-parallel, so that the lateral beam displacements they introduce cancel each other and the alignment is preserved when the compensation is changed. The insertion loss of the compensation stage is only $0.4\%$, which is the main practical advantage of the Brewster-angle free-space implementation over the spliced dispersion-compensating fiber of Ref.~\cite{Takanashi2020}.

The output is measured using an optical spectrum analyzer (OSA), which provides a direct measurement of the optical power spectrum.
The spectra were recorded with a Yokogawa AQ6380 over $184$--$204~\mathrm{THz}$ ($194\pm10~\mathrm{THz}$) at a resolution of $0.1~\mathrm{nm}$ ($12.55~\mathrm{GHz}$ at the carrier); resolution correction was enabled, so that the recorded traces are true power spectral densities. They were acquired with $15975$ sampling points at $0.01~\mathrm{nm}$ spacing ($1.255~\mathrm{GHz}$ at the carrier), in a single sweep without averaging or smoothing.

The shot-noise level is obtained by blocking the pump of OPA$_{\mathrm{sqz}}$, so that a vacuum state is injected into OPA$_{\mathrm{meas}}$, and recording the resulting amplified-vacuum spectrum with the gain, temperature and lock conditions of OPA$_{\mathrm{meas}}$ identical to those of the squeezing measurement. This trace therefore constitutes the correct standard-quantum-limit reference for the PSA measurement. Its spectral shape is not flat because it reflects the gain profile of OPA$_{\mathrm{meas}}$; in principle the same trace can also be used to determine the group-delay dispersion of the measurement waveguide, since its shape is governed by $r_\mathrm{meas.}$ and $D_\mathrm{meas.}$ through the Bloch--Messiah parameters of Sec.~2.3. 

\begin{table}[htbp]
\centering
\caption{Loss budget of the squeezing measurement.}
\label{tab:loss-budget}
\begin{tabular}{lc}
\hline\hline
Contribution & Efficiency \\
\hline
OPA$_{\mathrm{sqz}}$ waveguide (effective value at $13.2~\mathrm{dB}$ gain) & $0.974^{*}$ \\
Free-space propagation between the two OPAs & $0.950$ \\
Dispersion-compensation plates (Brewster angle) & $0.996$ \\
Coupling from free space into the OPA$_{\mathrm{meas}}$ waveguide & $0.872$ \\
OPA$_{\mathrm{meas}}$ waveguide (effective value at $25.3~\mathrm{dB}$ gain) & $0.982^{*}$ \\
Residual phase fluctuation & negligible \\
\hline
Total detection efficiency & $0.79$ \\
\hline\hline
\end{tabular}
\par\vspace{4pt}
\begin{minipage}{0.92\linewidth}
\footnotesize
$^{*}$These are estimated values.
\end{minipage}
\end{table}
The loss budget of the setup is summarized in Table~\ref{tab:loss-budget}. The efficiencies of the two OPAs are estimated from the loss of waveguides of the same structure together with the parametric gains used here, following the distributed-loss analysis of Appendix~\ref{app:lossy}. Taking all contributions into account, the total effective detection efficiency is $\eta=0.79$, which reproduces the measured squeezing level to within $0.2~\mathrm{dB}$ (Sec.~3.2).

The experimental results obtained for different thicknesses of the fused silica plates are shown in Fig.~\ref{fig:experimental_result}, corresponding to the simulation results discussed in the previous section. As explained above, the dispersion compensation cannot be continuously tuned in the present setup and therefore does not exactly match the optimal value derived in the previous section. However, by fine-tuning the pump power of the OPA, the effective quadrature rotation can be adjusted, enabling broadband PSA measurement, as demonstrated in the figure. From a practical standpoint, the absolute values of the squeezing level and the PSA gain are often not very critical. Therefore, adjusting the effective rotation via the pump power provides a more flexible and practical approach than implementing continuously tunable dispersion compensation.

\subsection{Experimental results}
\begin{figure}[htbp]  
\centering  
\includegraphics[width=13cm]{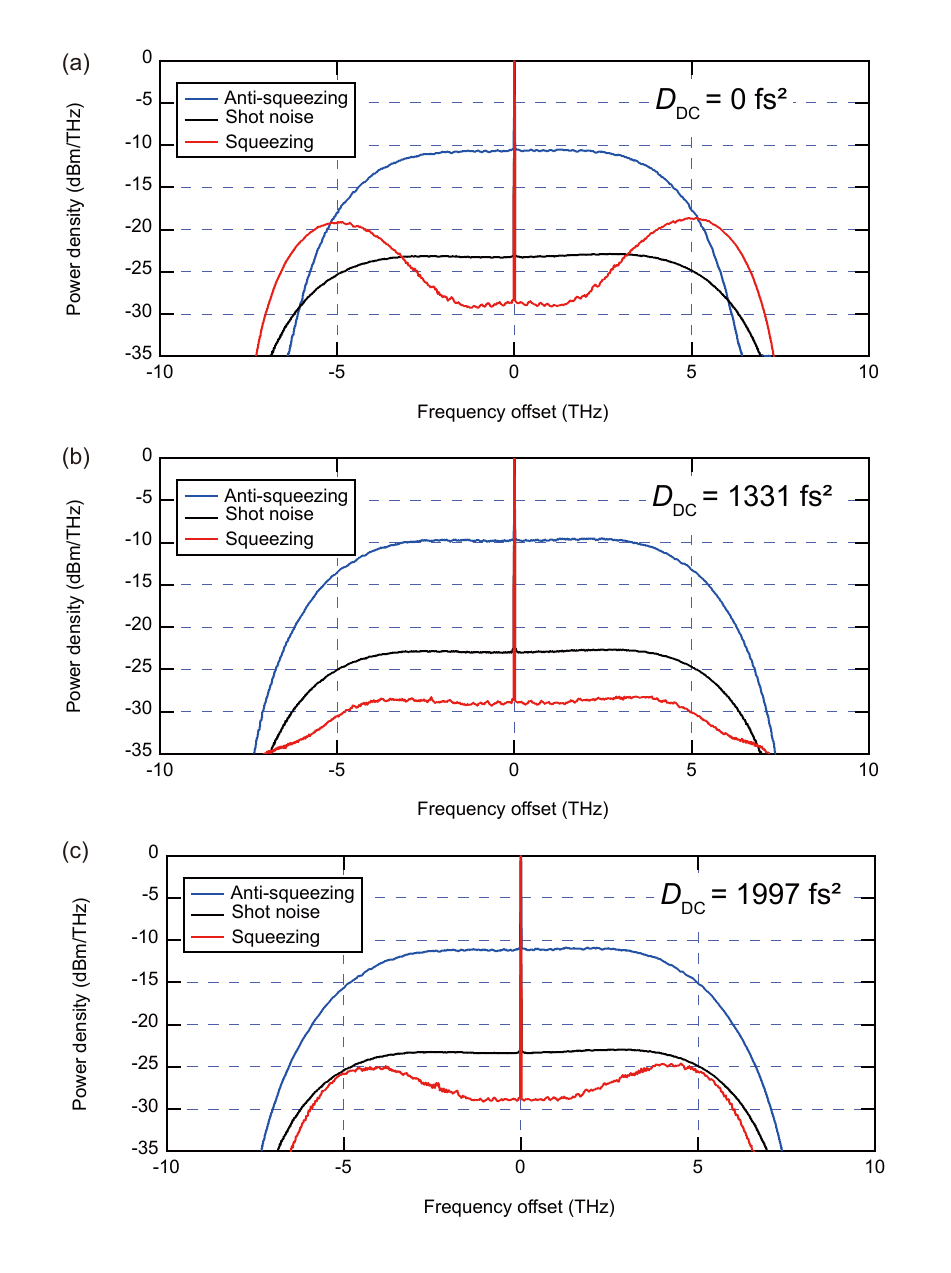}
\caption{Measured spectra of squeezing (red), anti-squeezing (blue), and shot noise (black) for (a) $D_{\mathrm{DC}}=0$, (b) $1331\pm8$, and (c) $1997\pm12$~$\mathrm{fs^2}$ (overcompensation), where $D_{\mathrm{DC}}\equiv|D_{\mathrm{dc}}|$. Panel (b), the accessible compensation closest to the optimum, gives the broadest squeezed band.}\label{fig:experimental_result}
\end{figure}

Figure~\ref{fig:experimental_result} shows the measured optical spectra obtained while varying the amount of dispersion compensation. In each panel, the red, blue, and black traces correspond to the squeezing spectrum, anti-squeezing spectrum, and shot-noise spectrum, respectively. 
Figure~\ref{fig:experimental_result}(a) shows the result obtained without the fused-silica plates used for dispersion compensation. Although approximately \SI{6}{dB} of squeezing relative to the shot-noise level is observed within about \SI{1.4}{THz} from the carrier frequency, the squeezing rapidly degrades at larger frequency offsets. As discussed in the theoretical analysis and numerical simulations, this degradation originates from the frequency-dependent quadrature rotation induced by residual dispersion in the two-stage OPA system.

Figure~\ref{fig:experimental_result}(b) shows the result obtained when dispersion close to the optimal compensation value was introduced using a pair of fused-silica plates. Compared with Fig.~\ref{fig:experimental_result}(a), squeezing is observed over a significantly broader frequency range. A maximum squeezing of $5.9\pm0.2~\mathrm{dB}$ relative to the shot-noise level is obtained near the carrier frequency, together with an anti-squeezing level of $13.2\pm0.1~\mathrm{dB}$. More specifically, the noise power remains more than \SI{5}{dB} below the shot-noise level even at frequency offsets as large as $5.9\pm0.1~\mathrm{THz}$ from the carrier frequency, and squeezing below the shot-noise level is maintained up to $6.9\pm0.1~\mathrm{THz}$. These results demonstrate that appropriate dispersion compensation effectively suppresses the frequency-dependent quadrature rotation and substantially broadens the measurable squeezing bandwidth.
The quoted levels are averaged over a residual ripple of about
$0.25~\mathrm{dB}$ present on the squeezing trace; this ripple dominates their
uncertainty, the statistical contribution being negligible, and the bandwidth
uncertainties follow from the local slope of the spectrum. The measured level agrees with the $6.0~\mathrm{dB}$ predicted in Sec.~2.5 with $\eta=0.79$.

Figure~\ref{fig:experimental_result}(c) shows the case of excessive dispersion compensation beyond the optimal value. When the compensation exceeds the optimum, the squeezing level at higher frequency offsets again becomes degraded. This behavior is also consistent with the theoretical prediction that overcompensation introduces an opposite quadrature rotation at higher frequencies.

The present results experimentally demonstrate that broadband squeezing measurement over several terahertz can be realized using a simple and low-loss dispersion-compensation scheme. In the present experiment, the dispersion compensation was adjusted to a value close to the optimum using fused-silica plates. However, as discussed above, it is not always possible to realize the exact optimal dispersion value because only discrete thicknesses of optical components are available. Even in such cases, as discussed in Fig.~\ref{fig.simulation_22mm_45mm}, the measurable bandwidth can still be broadened by slightly adjusting the pump powers of the two OPAs, thereby modifying the effective group-delay dispersion accumulated in each OPA.

\section{Conclusion}
OPAs are key devices for ultrafast optical quantum information processing, enabling the generation of squeezed vacuum states with bandwidths extending into the terahertz regime as well as their ultrafast measurement. However, in practical implementations, GVD inside the OPA induces a frequency-dependent rotation of the squeezing axis, which restricts the measurable squeezing up to a frequency offset of 1 THz, even when the phase-matching bandwidth is significantly broader.

In this work, we have demonstrated that this limitation can be substantially overcome by introducing external dispersion compensation in a simple and low-loss two-stage waveguide OPA system. As a result, we achieved a maximum squeezing of $5.9\pm0.2$ dB near the carrier frequency, maintained more than 5 dB of squeezing up to a frequency offset of $5.9\pm0.1$ THz from the carrier, and observed squeezing below the shot-noise level up to an offset of $6.9\pm0.1$ THz. The measured squeezing level is quantitatively accounted for by the measured total detection efficiency of $0.79$, the dominant contribution being the coupling of the squeezed field into the measurement waveguide.

These results show that the measurable bandwidth can be extended by several times compared to conventional approaches, enabling efficient and parallel characterization of broadband squeezed light. The theoretical framework developed here provides a practical design principle for dispersion engineering in cascaded parametric systems, and the demonstrated scheme offers a scalable route toward ultrafast continuous-variable quantum information processing.

\begin{backmatter}
\bmsection{Funding}
A part of this work was supported by Japan Science and Technology Agency (JST) Moonshot R\&D (Grant No. JPMJMS2064 and JPMJMS256I), and Council for Science, Technology and Innovation (CSTI), Cross-ministerial Strategic Innovation Promotion Program (SIP), ``Promoting Application of Advanced Quantum Technologies to Social Challenges'' (Project management agency: QST). T.S. acknowledges financial support from The Forefront Physics and Mathematics Program to Drive Transformation (FoPM), a World-leading Innovative Graduate Study (WINGS) Program. T.S. acknowledges support from Japan Society for the Promotion of Science (JSPS) Research Fellowship for Young Scientists (No. 25KJ0768). M.E. acknowledges funding from JST (JPMJPR2254) and JSPS KAKENHI (No. 24K01374).

\bmsection{Acknowledgment}
The authors acknowledge support from UTokyo Foundation and donations from Nichia Corporation of Japan.

\bmsection{Disclosures}
The authors declare no conflicts of interest.

\bmsection{Data Availability Statement}
Data underlying the results presented in this paper are not publicly available at this time but may be obtained from the authors upon reasonable request.

\bmsection{Author contributions}
T.S.\ proposed the method and conceived the study. Under the guidance of T.S.,
S.O.\ built the experimental setup and carried out the theoretical modeling,
analysis, and numerical simulations. T.S., K.H., T.H., R.I., and M.E.\ provided
technical and theoretical supports for the study. T.K., T.Y., A.I., and T.U.\ fabricated
and characterized the PPLN waveguides. A.F.\ supervised the project. T.S., 
and M.E.\ wrote the manuscript with support from S.O. and all co-authors.
\end{backmatter}

\appendix
\makeatletter
\renewcommand{\@seccntformat}[1]{Appendix~\csname the#1\endcsname:\quad}
\makeatother

\section{Bloch--Messiah decomposition of $A(r,\theta)$}
\label{app:bloch-messiah}

In this Appendix, we derive the explicit form of the matrix
$A(r,\theta)$ defined in Sec.~2.3 of the main text and obtain its
Bloch--Messiah decomposition in all parameter regimes.

\subsection{Matrix exponential of the generator}
\label{app:matrix-exp}

The simultaneous action of squeezing and dispersion inside the
waveguide is described by $A(r,\theta)=e^{B}$ with the traceless
generator
\begin{equation}
  B \;=\;
  \begin{pmatrix}
    r & -\theta \\
    \theta & -r
  \end{pmatrix}.
  \label{eq:app-B}
\end{equation}
A direct calculation gives
\begin{equation}
  B^{2} \;=\; (r^{2}-\theta^{2})\,I,
  \label{eq:app-Bsquared}
\end{equation}
so that every power of $B$ reduces to a linear combination of $I$
and $B$,
\begin{equation}
  B^{2n} \;=\; (r^{2}-\theta^{2})^{n}\,I,
  \qquad
  B^{2n+1} \;=\; (r^{2}-\theta^{2})^{n}\,B.
  \label{eq:app-Bpowers}
\end{equation}

Separating the exponential series into its even and odd parts and
introducing
\begin{equation}
  k \;\equiv\; \sqrt{\,\lvert r^{2}-\theta^{2}\rvert\,},
  \label{eq:app-k}
\end{equation}
we obtain a closed-form expression for $A$ in three regimes:
\begin{equation}
  A(r,\theta) \;=\;e^B \;=\;
  \begin{cases}
    \displaystyle
    I\cosh k \;+\; \frac{\sinh k}{k}\,B
    & (r^{2}>\theta^{2}),\\[10pt]
    \displaystyle
    I \;+\; B
    & (r^{2}=\theta^{2}),\\[6pt]
    \displaystyle
    I\cos k \;+\; \frac{\sin k}{k}\,B
    & (r^{2}<\theta^{2}).
  \end{cases}
  \label{eq:app-A-three-cases}
\end{equation}

Writing out the components explicitly,
\begin{equation}
  A(r,\theta) \;=\;
  \begin{pmatrix}
    \cosh k + \dfrac{r}{k}\sinh k &
    -\dfrac{\theta}{k}\sinh k \\[10pt]
    \dfrac{\theta}{k}\sinh k &
    \cosh k - \dfrac{r}{k}\sinh k
  \end{pmatrix}
  \qquad (r^{2}>\theta^{2}),
  \label{eq:app-A-hyperbolic}
\end{equation}
\begin{equation}
  A(r,\theta) \;=\;
  \begin{pmatrix}
    1+r & -\theta \\
    \theta & 1-r
  \end{pmatrix}
  \qquad (r^{2}=\theta^{2}),
  \label{eq:app-A-shear}
\end{equation}
\begin{equation}
  A(r,\theta) \;=\;
  \begin{pmatrix}
    \cos k + \dfrac{r}{k}\sin k &
    -\dfrac{\theta}{k}\sin k \\[10pt]
    \dfrac{\theta}{k}\sin k &
    \cos k - \dfrac{r}{k}\sin k
  \end{pmatrix}
  \qquad (r^{2}<\theta^{2}).
  \label{eq:app-A-trigonometric}
\end{equation}

\subsection{Bloch--Messiah decomposition}
\label{app:bm-decomp}

The Bloch--Messiah decomposition
\begin{equation}
  A(r,\theta) \;=\; R(\phi)\,S(r')\,R(\phi'),
  \label{eq:app-bm}
\end{equation}
where $R(\cdot)$ and $S(\cdot)$ are the rotation and squeezing matrices defined in the main text, is summarized in Table~\ref{tab:bm-regimes}. Each entry can be verified by direct substitution into Eq.~\eqref{eq:app-bm}.

\begin{table}[h]
  \centering
  \caption{Bloch--Messiah decomposition $A(r,\theta)=R(\phi)S(r')R(\phi')$ in the three regimes $r\gtrless\theta$, with $k=\sqrt{|r^2-\theta^2|}$. The boundary $r=\theta$ occurs at a frequency offset of $\sim5$--$6~\mathrm{THz}$ from the carrier here.}
  \label{tab:bm-regimes}
  \begin{tabular}{cccc}
    \hline\hline
    Regime & $k$ & $r'$ & $\phi=\phi'$ \\
    \hline
    $r^{2}>\theta^{2}$ &
    $\sqrt{r^{2}-\theta^{2}}$ &
    $\operatorname{arcsinh}\left(\tfrac{r}{k}\sinh k\right)$ &
    $\tfrac{1}{2}\arctan\left(\tfrac{\theta}{k}\tanh k\right)$ \\[4pt]
    $r^{2}=\theta^{2}$ &
    $0$ &
    $\operatorname{arcsinh}(r)$ &
    $\tfrac{1}{2}\arctan(r)$ \\[4pt]
    $r^{2}<\theta^{2}$ &
    $\sqrt{\theta^{2}-r^{2}}$ &
    $\operatorname{arcsinh}\left(\tfrac{r}{k}\sin k\right)$ &
    $\tfrac{1}{2}\arctan\left(\tfrac{\theta}{k}\tan k\right)^{\dagger}$ \\
    \hline\hline
  \end{tabular}
  \par\vspace{4pt}
  \begin{minipage}{0.92\linewidth}
    \footnotesize
    $^{\dagger}$In the trigonometric regime $r^{2}<\theta^{2}$, the arctangent is its principal value only for $k<\pi/2$; beyond, it is continued smoothly across the poles of $\tan k$ at $k=(n+\tfrac{1}{2})\pi$ as $\phi=\tfrac{1}{2}\!\left[\arctan\!\left(\tfrac{\theta}{k}\tan k\right)+m\pi\right]$ with $m=\left\lfloor k/\pi+\tfrac{1}{2}\right\rfloor$. In turn, $r'=\operatorname{arcsinh}\!\left(\tfrac{r}{k}\sin k\right)$ changes sign each time $k$ crosses a multiple of $\pi$, where the squeezed and anti-squeezed quadratures interchange.
  \end{minipage}
\end{table}

The first row of Table~\ref{tab:bm-regimes} reproduces the analytic expressions for $a$ and $\alpha=\beta$ given in Sec.~2.3 of the main text (with the identification $r'\leftrightarrow a$ and $\phi\leftrightarrow\alpha$). 
Each entry can be verified by direct substitution, the rotation angle in the trigonometric regime being continued as specified in the table footnote.

\subsection{Sideband degradation of the effective squeezing
parameter $r'$}
\label{app:gain-degradation}

The Bloch--Messiah decomposition of Sec.~\ref{app:bm-decomp}
yields not only a frequency-dependent rotation $\phi(\Omega)$
but also a frequency-dependent effective squeezing parameter
$r'(\Omega)=\operatorname{arcsinh}\bigl(\tfrac{r}{k}\sinh k\bigr)$,
with $k=\sqrt{r^{2}-\theta^{2}}$. Whereas $\phi$ can in
principle be cancelled by external dispersion compensation, the
deviation of $r'$ from $r$ is intrinsic to the waveguide and
cannot be undone downstream. In this subsection, we quantify this
intrinsic degradation for small frequency offsets $\theta\ll r$.

Expanding $r'(\theta)$ for small $\theta$, we obtain
\begin{equation}
  r'(\theta) \;\simeq\; r \;-
  \frac{\theta^{2}}{2r}
  \left( 1 - \frac{\tanh r}{r} \right).
  \label{eq:app-gain-drop}
\end{equation}

Substituting $\theta(\Omega)=D\Omega^{2}/2$ into
Eq.~\eqref{eq:app-gain-drop}, the effective squeezing parameter becomes
\begin{equation}
  r'(\Omega) \;\simeq\; r
  \;-
  \frac{D^{2}\Omega^{4}}{8r}
  \left(1-\frac{\tanh r}{r}\right).
  \label{eq:app-gain-degradation-Omega}
\end{equation}
Equation~\eqref{eq:app-gain-degradation-Omega} shows that the
deviation of $r'(\Omega)$ from $r$ scales as $\mathcal{O}(\Omega^{4})$.
This quartic dependence indicates that the parametric-gain spectrum
is remarkably flat-topped around the carrier frequency. This is in
contrast to the phase-rotation angle $\phi(\Omega)$, whose leading
frequency dependence is $\mathcal{O}(\Omega^{2})$. The $\Omega^{4}$
scaling therefore helps explain why optical parametric amplifiers
can preserve ultrabroadband squeezing despite dispersion.

\section{Lossy Squeezer and Optimal Dispersion Compensation}
\label{app:lossy}

In this Appendix, we extend the lossless analysis of Sec.~2 to
include intrinsic loss in the squeezer. In the lossy case, vacuum
noise is continuously admixed from the environment, so the same
strategy as in the lossless case is no longer convenient.
Fortunately, because the states considered here remain Gaussian,
the problem can still be analyzed efficiently by tracking the
covariance matrix. We derive the covariance matrix of the output
field in the continuum limit and show, both qualitatively and
numerically, that the optimal dispersion compensation decreases as
loss increases.

\subsection{Model of lossy squeezer}
\label{app:lossy-model}

Following the same sideband picture as in the main text, we model
a lossy squeezer by dividing the waveguide into $N$ slices, each
consisting of an infinitesimal squeezing--rotation operation
$R(\theta/N)\,S(r/N)$ followed by a beam splitter of transmittance
$T^{1/N}$ that mixes the field with an environmental vacuum. In
the limit $N\to\infty$, this reproduces a uniformly distributed
loss along the waveguide with total transmittance $T$.

Since the quadrature means vanish throughout, we work directly
with the covariance matrix. Denoting by $M_n$ the covariance
matrix after $n$ slices and by $M_{\mathrm{vac}}=I$ the vacuum
covariance matrix, the recursion reads
\begin{equation}
  M_{n+1}
  \;=\;
  T^{1/N} R(\theta/N)\,S(r/N)\,M_n\,S^{T}(r/N)\,R^{T}(\theta/N)
  \;+\;(1-T^{1/N})\,M_{\mathrm{vac}}.
  \label{eq:app-recursion}
\end{equation}

Defining
\begin{align}
  \mathcal{A}_N &\;\equiv\; T^{1/(2N)}\,R(\theta/N)\,S(r/N),
  \label{eq:app-AN}\\
  \mathcal{B}_N &\;\equiv\; (1-T^{1/N})\,M_{\mathrm{vac}},
  \label{eq:app-BN}
\end{align}
the recursion becomes
$M_{n+1}=\mathcal{A}_N M_n \mathcal{A}_N^{T}+\mathcal{B}_N$.
Introducing the fixed-point matrix $Z$ satisfying
\begin{equation}
  Z \;=\; \mathcal{A}_N\,Z\,\mathcal{A}_N^{T} \;+\; \mathcal{B}_N,
  \label{eq:app-Z-def}
\end{equation}
the shifted quantity $M_n-Z$ evolves homogeneously and the solution
after $N$ slices is
\begin{equation}
  M_N \;=\; (\mathcal{A}_N)^{N}\,(M_0-Z)\,(\mathcal{A}_N^{T})^{N}
  \;+\; Z.
  \label{eq:app-MN-solution}
\end{equation}

\subsection{Continuum limit and Lyapunov equation}
\label{app:lyapunov}

Taking $N\to\infty$ and keeping terms up to $\mathcal{O}(1/N)$,
\begin{align}
  \mathcal{A}_N &\simeq
  \left(1+\tfrac{\log T}{2N}\right)\!\left(I+\tfrac{B}{N}\right),
  \label{eq:app-AN-expansion}\\
  \mathcal{B}_N &\simeq -\tfrac{\log T}{N}\,M_{\mathrm{vac}},
  \label{eq:app-BN-expansion}
\end{align}
where $B$ is the generator defined in Sec.~2.3 of the main text.
Substituting into Eq.~\eqref{eq:app-Z-def} and collecting
$\mathcal{O}(1/N)$ terms yields the Lyapunov equation
\begin{equation}
  \log T\,(Z-I) \;+\; B\,Z \;+\; Z\,B^{T} \;=\; 0.
  \label{eq:app-lyapunov}
\end{equation}
Writing $Z$ as a symmetric $2\times 2$ matrix,
\begin{equation}
  Z \;=\; \begin{pmatrix} x & y \\ y & z \end{pmatrix},
  \label{eq:app-Z-form}
\end{equation}
and defining $\lambda\equiv \log T$, the entry-wise solution of
Eq.~\eqref{eq:app-lyapunov} is
\begin{equation}
  x \;=\; \frac{\lambda^{2}-2\lambda r+4\theta^{2}}
               {\lambda^{2}+4\theta^{2}-4r^{2}},
  \quad
  y \;=\; \frac{4\theta r}{\lambda^{2}+4\theta^{2}-4r^{2}},
  \quad
  z \;=\; \frac{\lambda^{2}+2\lambda r+4\theta^{2}}
               {\lambda^{2}+4\theta^{2}-4r^{2}}.
  \label{eq:app-Z-solution}
\end{equation}

The output covariance matrix in the continuum limit is obtained by
letting $(\mathcal{A}_N)^{N}\to \sqrt{T}\,e^{B}$, giving
\begin{equation}
  M(M_0,T,r,\theta) \;=\;
  T\,e^{B}\,M_0\,(e^{B})^{T}
  \;-\; T\,e^{B}\,Z\,(e^{B})^{T}
  \;+\; Z,
  \label{eq:app-M-final}
\end{equation}
where the first term describes the lossy squeezing of the input
state and the remaining two terms represent the vacuum
contamination from the distributed loss.

\subsection{Analytic diagonalization of the output from a single squeezer}
\label{app:diag}

For a vacuum input $M_0=I$, Eq.~\eqref{eq:app-M-final} gives
\begin{equation}
  M(I,T,r,\theta)
  \;=\;
  T\,e^{B}(e^{B})^{T}
  \;-\; T\,e^{B}Z(e^{B})^{T}
  \;+\; Z.
  \label{eq:app-M-vacuum}
\end{equation}
Since this output covariance matrix is a real symmetric
positive-definite $2\times 2$ matrix, it can be brought to the
symplectic standard form (Williamson decomposition):
\begin{equation}
  M \;=\; \nu\,R(\varphi)\,S(2\tilde a)\,R(\varphi)^{T},
  \label{eq:app-diag-form}
\end{equation}
where $\nu\ge 1$ is the thermal impurity, $\tilde a$ is the symplectic squeezing parameter,
and $\varphi$ is the orientation of the squeezed quadrature in phase space.
These parameters are extracted from the matrix elements via:
\begin{equation}
  \nu \;=\; \sqrt{\det M},
  \qquad
  \cosh 2\tilde a \;=\; \frac{\operatorname{tr} M}{2\nu},
  \qquad
  \tan 2\varphi \;=\; \frac{2M_{12}}{M_{11}-M_{22}}.
  \label{eq:app-diag-extract}
\end{equation}

\paragraph{Matrix elements and rotation angle.}
Using Eq.~\eqref{eq:app-M-vacuum} and the solution for $Z$, the
matrix elements entering Eq.~\eqref{eq:app-diag-extract} for $M_0=I$ are:
\begin{align}
  M_{11}-M_{22} &\;=\; \frac{4r}{\Delta}\,
    \bigl[\,\lambda\,(T\cosh 2k-1)-2Tk\sinh 2k\,\bigr],
    \label{eq:app-M11-M22}\\[2pt]
  M_{12} &\;=\; \frac{4\theta r}{\Delta}\,
    \Bigl[\,\tfrac{T\lambda}{2k}\sinh 2k + 1-T\cosh 2k\,\Bigr],
    \label{eq:app-M12}
\end{align}
with $k=\sqrt{r^{2}-\theta^{2}}$, $\lambda=\log T$, and
$\Delta=\lambda^{2}+4\theta^{2}-4r^{2}$. Here we assume the regime $r^{2}>\theta^{2}$. 
Substituting these into 
Eq.~\eqref{eq:app-diag-extract} yields the exact expression:
\begin{equation}
  \tan 2\varphi \;=\;
  \frac{\theta\,(2kX-\lambda Y)}{k\,(2kY-\lambda X)},
  \qquad
  X \equiv T\cosh 2k - 1,
  \quad
  Y \equiv T\sinh 2k.
  \label{eq:app-varphi-closed}
\end{equation}

\paragraph{Loss effect on effective GDD.}
To find the effective dispersion, we expand $\varphi(\theta)$ around
the carrier frequency ($\theta \to 0$). We obtain
$\varphi(\Omega)\simeq \kappa(r,T)\,\theta(\Omega)$ with:
\begin{equation}
  \kappa(r,T) \;=\;
  \frac{1}{2r}\,\frac{2rX_{0}-\lambda Y_{0}}
                      {2rY_{0}-\lambda X_{0}},
  \qquad
  X_{0} \equiv T\cosh 2r - 1,
  \quad
  Y_{0} \equiv T\sinh 2r.
  \label{eq:app-kappa}
\end{equation}
Here $D$ denotes the physical group-delay dispersion accumulated
in the waveguide, as introduced through $\theta(\Omega)=D\Omega^2/2$.
The effective group-delay dispersion $D_{\mathrm{eff}}$ experienced by
the output squeezed state is then related to $D$ by
\begin{equation}
  D_{\mathrm{eff}}(r,T) \;=\; \kappa(r,T)\,D.
\end{equation}
In the lossless limit $T \to 1$, $\lambda \to 0$, $X_0 \to \cosh 2r-1$, 
and $Y_0 \to \sinh 2r$, this reduces to:
\begin{equation}
  \kappa(r,1) \;=\; \frac{1}{2r}\frac{\cosh 2r-1}{\sinh 2r} 
  \;=\; \frac{\tanh r}{2r},
\end{equation}
recovering the lossless scaling factor.

\paragraph{Proof of monotonicity in $T$.}
We prove that $\kappa(r,T)$ is monotonically increasing in $T$ on
$(0,1]$ for every $r>0$. Introducing $\lambda=\log T$ and
dividing the numerator and denominator of
Eq.~\eqref{eq:app-varphi-closed} evaluated at $\theta\to 0$ by $T$,
\begin{equation}
  \kappa(r,\lambda) \;=\;
  \frac{1}{2r}\,
  \frac{P(\lambda)}{Q(\lambda)},
  \qquad
  \begin{aligned}
    P(\lambda) &\equiv 2r(c-e^{-\lambda})-\lambda s,\\
    Q(\lambda) &\equiv 2rs-\lambda(c-e^{-\lambda}),
  \end{aligned}
  \label{eq:app-kappa-lambda}
\end{equation}
with $c\equiv\cosh 2r$ and $s\equiv\sinh 2r$. Since
$\partial\kappa/\partial T$ and $\partial\kappa/\partial\lambda$
share the same sign for $T>0$, it suffices to show that
$\kappa$ is monotonically non-decreasing in $\lambda$.

\medskip

Using the identity $e^{2r}=c+s$, one verifies
$P(-2r)=Q(-2r)=0$, so that $\lambda=-2r$ is a removable
singularity of $P/Q$: both $P$ and $Q$ vanish to first order in
$(\lambda+2r)$, and their ratio extends continuously to a finite
value. Away from this point, $Q(\lambda)\neq 0$, and direct
differentiation of Eq.~\eqref{eq:app-kappa-lambda} together with
$c^{2}-s^{2}=1$ gives
\begin{equation}
  \frac{\partial\kappa}{\partial\lambda}
  \;=\;
  \frac{1}{2r}\,\frac{e^{-\lambda}\,G(\lambda,r)}{Q(\lambda)^{2}}
  \qquad (\lambda\neq -2r),
  \label{eq:app-dkappa-dlambda}
\end{equation}
with
\begin{equation}
  G(\lambda,r) \;\equiv\;
  4r\,(\cosh\lambda - \cosh 2r)
  \;+\;
  (4r^{2}-\lambda^{2})\sinh 2r.
  \label{eq:app-G-def}
\end{equation}

\medskip

We show $G(\lambda,r)\ge 0$. The function is even in $\lambda$;
with $\ell\equiv|\lambda|\ge 0$,
\begin{equation}
  \frac{\partial G}{\partial\ell}
  \;=\;
  2\ell\!\left(\,2r\,\frac{\sinh\ell}{\ell}-\sinh 2r\,\right).
  \label{eq:app-dG-dl}
\end{equation}
Since $x\mapsto\sinh x/x$ is strictly increasing on
$(0,\infty)$, the bracket vanishes only at $\ell=2r$, is negative
for $0<\ell<2r$, and positive for $\ell>2r$. Hence $G$ attains
its global minimum on $\ell\ge 0$ at $\ell=2r$, with
\begin{equation}
  G(\pm 2r,r) \;=\; 0,
  \label{eq:app-G-min}
\end{equation}
so that $G(\lambda,r)\ge 0$ for all $\lambda\in\mathbb{R}$.

\medskip

Combining Eqs.~\eqref{eq:app-dkappa-dlambda}
and~\eqref{eq:app-G-min}, we obtain
$\partial\kappa/\partial\lambda\ge 0$ wherever the derivative is
well defined. Thus $\kappa$ is non-decreasing as a function of
$\lambda$. The point $\lambda=-2r$ is only a removable singularity,
so this monotonicity extends across it by continuity. Since the
physical domain corresponds to $\lambda=\log T\le 0$ (hence the point
$\lambda=+2r$ is irrelevant here), we conclude that
$\kappa(r,T)$ is monotonically non-decreasing in $T$ on $(0,1]$.

\begin{figure}[t]
  \centering
  \includegraphics[width=\linewidth]{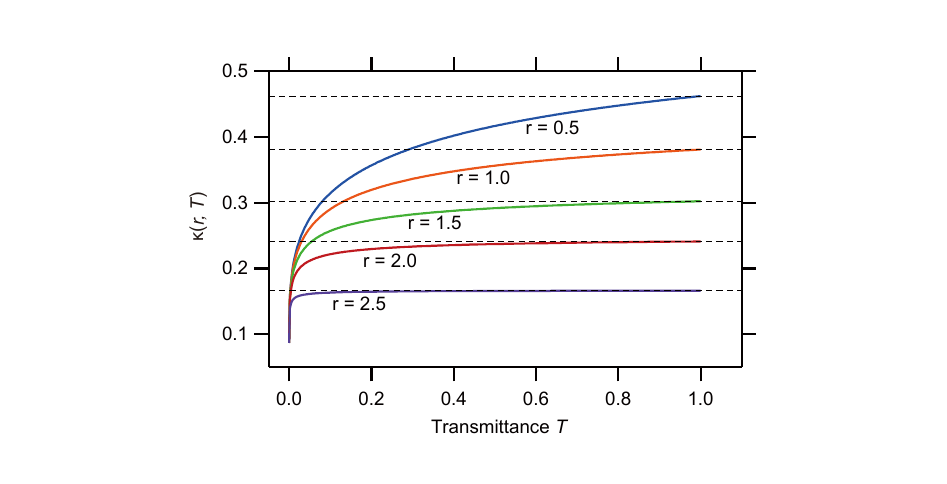}
  \caption{Dependence of $\kappa(r,T)$ on the waveguide transmittance $T$. The monotonic increase of $\kappa$ with $T$ implies that stronger loss leads to a smaller effective dispersion parameter $D_{\mathrm{eff}}=\kappa D$.}
  \label{fig:kappa-monotonicity}
\end{figure}

\paragraph{Physical interpretation.}
Because $D_{\mathrm{eff}}(r,T)=\kappa(r,T)D$, the monotonic increase of
$\kappa$ with $T$ means that the effective GDD decreases as the loss
in the waveguide increases. Intuitively, stronger loss implies a
larger contribution from vacuum modes admixed partway along the
waveguide. These vacuum modes do not experience the full phase
rotation accumulated from the input facet, and therefore carry a
smaller phase rotation than vacuum fluctuations injected from the
beginning of the propagation. As their weight increases, the net
phase rotation of the output squeezed state is reduced, leading to
a smaller effective GDD.

\subsection{All-optical measurement and effective efficiency}
\label{app:meas-eff}

In an all-optical measurement setup, the performance of the lossy squeezer is evaluated by directly measuring the intensity (energy) of the output field. For a zero-mean Gaussian state, the average photon number at frequency $\Omega$ is determined by the trace of the covariance matrix as $\langle \hat{n}(\Omega) \rangle = \frac{1}{4}(\operatorname{tr} M(\Omega) - 2)$.

For an arbitrary input state characterized by $M_0$, substituting Eq.~\eqref{eq:app-M-final} into this relation yields the total output energy. The trace of the output covariance matrix naturally splits into two components:
\begin{equation}
  \operatorname{tr} M
  \;=\;
  \operatorname{tr} \bigl[\, T e^B M_0 (e^B)^T \,\bigr]
  \;+\;
  \operatorname{tr} N_{\mathrm{cont}},
  \qquad
  N_{\mathrm{cont}} \;\equiv\; Z - T e^B Z (e^B)^T.
\end{equation}
The first term represents the pure lossy amplification of the input state $M_0$. The second term, $N_{\mathrm{cont}}$, represents the excess noise strictly generated by the loss-induced environmental vacuum fields continuously admixed along the waveguide.

To interpret this result practically, we consider the all-optical measurement in the strong-amplification regime ($e^{2r'} \gg 1$). Suppose that the input carries a signal in the amplified quadrature with variance $\langle x_{\phi}^{2} \rangle$. After propagation, the purely amplified signal contribution scales as:
\begin{equation}
 \operatorname{tr} \bigl[\, T e^B M_0 (e^B)^T \,\bigr]
  \;\simeq\;
  T e^{2r'}\,\langle x_{\phi}^{2} \rangle.
\end{equation}
Unlike the direct measurement of the squeezed vacuum discussed earlier---where distributed loss slightly modifies the optimal squeezing angle---the rotation of the measurement phase in this all-optical scheme is completely unaffected by the optical loss inside the waveguide.

The exact trace of the added-noise term can be written analytically as:
\begin{equation}
  \operatorname{tr} N_{\mathrm{cont}}
  \;=\;
  \frac{2}{\Delta}
  \left[
    (\lambda^2+4\theta^2)(1-T)
    - \frac{Tr^2\lambda^2}{k^2}(\cosh 2k-1)
    + \frac{2T\lambda r^2}{k}\sinh 2k
  \right].
\end{equation}

Consequently, the average output photon number is approximated by the sum of these two contributions:
\begin{equation}
  \langle \hat{n} \rangle
  \;\simeq\;
  \frac{1}{4} \left[\, T e^{2r'}\,\langle x_{\phi}^{2} \rangle
  \;+\; \operatorname{tr} N_{\mathrm{cont}} \,\right].
\end{equation}
The first term inside the bracket represents the detected signal amplified by the macroscopic effective gain, whereas the second term represents the inevitable noise floor introduced by the distributed loss. By comparing the pure amplification factor to the total amplification including this noise, the effective measurement efficiency is naturally defined as:
\begin{equation}
  \eta_{\mathrm{eff}}
  \;\equiv\;
  \frac{T e^{2r'}}
       {T e^{2r'} + \operatorname{tr} N_{\mathrm{cont}}}.\label{eq:eta-eff}
\end{equation}

\begin{figure}[H]
  \centering
  \includegraphics[width=\linewidth]{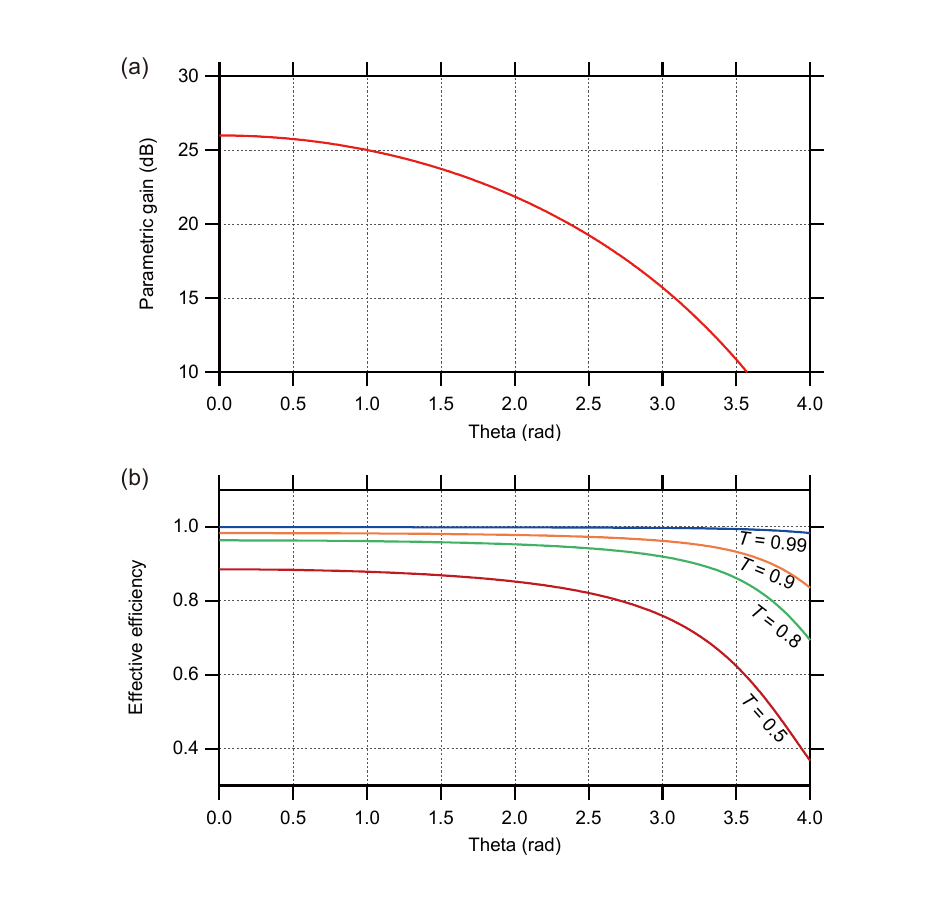}
  \caption{(a) Parametric gain $e^{2r'}$ and (b) effective measurement efficiency $\eta_{\mathrm{eff}}$ [Eq.~\eqref{eq:eta-eff}] as functions of the rotation parameter $\theta=D\Omega^2/2$, for a pump parameter $r=2.993$ (peak parametric gain $\approx 26$~dB). Curves correspond to waveguide transmittances $T=0.99,\,0.9,\,0.8,\,0.5$. Even when the effective gain drops to $\sim 15$~dB near the phase-matching boundary, $\eta_{\mathrm{eff}}$ remains high, demonstrating the robustness of the all-optical measurement against distributed loss.}
  \label{fig:efficiency-dispersion}
\end{figure}
\bibliography{DC}

\begin{thebibliography}{10}
\newcommand{\enquote}[1]{``#1''}

\bibitem{Braunstein2005}
S.~L. Braunstein and P.~v. Loock, \enquote{Quantum information with continuous
  variables,} {\protect\JournalTitle{Reviews of Modern Physics}} \textbf{77},
  513 (2005).

\bibitem{Giovannetti2004}
V.~Giovannetti, S.~Lloyd, and L.~Maccone, \enquote{Quantum-enhanced
  measurements: beating the standard quantum limit,}
  {\protect\JournalTitle{Science}} \textbf{306}, 1330--6 (2004).

\bibitem{Andersen2016}
U.~L. Andersen, T.~Gehring, C.~Marquardt, and G.~Leuchs, \enquote{30 years of
  squeezed light generation,} {\protect\JournalTitle{Physica Scripta}}
  \textbf{91}, 053001 (2016).

\bibitem{Furusawa1998}
A.~Furusawa, J.~L. Sorensen, S.~L. Braunstein, \emph{et~al.},
  \enquote{Unconditional quantum teleportation,}
  {\protect\JournalTitle{Science}} \textbf{282}, 706--9 (1998).

\bibitem{Braunstein1998}
S.~L. Braunstein and H.~J. Kimble, \enquote{Teleportation of continuous quantum
  variables,} {\protect\JournalTitle{Physical Review Letters}} \textbf{80},
  869--872 (1998).

\bibitem{Asavanant2019}
W.~Asavanant, Y.~Shiozawa, S.~Yokoyama, \emph{et~al.}, \enquote{Generation of
  time-domain-multiplexed two-dimensional cluster state,}
  {\protect\JournalTitle{Science}} \textbf{366}, 373--376 (2019).

\bibitem{Larsen2019}
M.~V. Larsen, X.~Guo, C.~R. Breum, \emph{et~al.}, \enquote{Deterministic
  generation of a two-dimensional cluster state,}
  {\protect\JournalTitle{Science}} \textbf{366}, 369--372 (2019).

\bibitem{Menicucci2006}
N.~C. Menicucci, P.~van Loock, M.~Gu, \emph{et~al.}, \enquote{Universal quantum
  computation with continuous-variable cluster states,}
  {\protect\JournalTitle{Phys Rev Lett}} \textbf{97}, 110501 (2006).

\bibitem{Raussendorf2001}
R.~Raussendorf and H.~J. Briegel, \enquote{A one-way quantum computer,}
  {\protect\JournalTitle{Phys Rev Lett}} \textbf{86}, 5188--91 (2001).

\bibitem{Raussendorf2003}
R.~Raussendorf, D.~E. Browne, and H.~J. Briegel, \enquote{Measurement-based
  quantum computation on cluster states,} {\protect\JournalTitle{Physical
  Review A}} \textbf{68}, 022312 (2003).

\bibitem{Vahlbruch2016}
H.~Vahlbruch, M.~Mehmet, K.~Danzmann, and R.~Schnabel, \enquote{Detection of 15
  db squeezed states of light and their application for the absolute
  calibration of photoelectric quantum efficiency,} {\protect\JournalTitle{Phys
  Rev Lett}} \textbf{117}, 110801 (2016).

\bibitem{Kashiwazaki2021}
T.~Kashiwazaki, T.~Yamashima, N.~Takanashi, \emph{et~al.}, \enquote{Fabrication
  of low-loss quasi-single-mode ppln waveguide and its application to a
  modularized broadband high-level squeezer,} {\protect\JournalTitle{Applied
  Physics Letters}} \textbf{119}, 251104 (2021).

\bibitem{Nehra2022}
R.~Nehra, R.~Sekine, L.~Ledezma, \emph{et~al.}, \enquote{Few-cycle vacuum
  squeezing in nanophotonics,} {\protect\JournalTitle{Science}} \textbf{377},
  1333--1337 (2022).

\bibitem{Kashiwazaki2020}
T.~Kashiwazaki, N.~Takanashi, T.~Yamashima, \emph{et~al.},
  \enquote{Continuous-wave 6-db-squeezed light with 2.5-thz-bandwidth from
  single-mode ppln waveguide,} {\protect\JournalTitle{APL Photonics}}
  \textbf{5}, 036104 (2020).

\bibitem{Shaked2018}
Y.~Shaked, Y.~Michael, R.~Z. Vered, \emph{et~al.}, \enquote{Lifting the
  bandwidth limit of optical homodyne measurement with broadband parametric
  amplification,} {\protect\JournalTitle{Nat Commun}} \textbf{9}, 609 (2018).

\bibitem{Kalash2023}
M.~Kalash and M.~V. Chekhova, \enquote{Wigner function tomography via optical
  parametric amplification,} {\protect\JournalTitle{Optica}} \textbf{10},
  1142--1146 (2023).

\bibitem{Takanashi2020}
N.~Takanashi, A.~Inoue, T.~Kashiwazaki, \emph{et~al.}, \enquote{All-optical
  phase-sensitive detection for ultra-fast quantum computation,}
  {\protect\JournalTitle{Opt Express}} \textbf{28}, 34916--34926 (2020).

\bibitem{Caves1982}
C.~M. Caves, \enquote{Quantum limits on noise in linear amplifiers,}
  {\protect\JournalTitle{Physical Review D}} \textbf{26}, 1817--1839 (1982).

\bibitem{Inoue2023}
A.~Inoue, T.~Kashiwazaki, T.~Yamashima, \emph{et~al.}, \enquote{Toward a
  multi-core ultra-fast optical quantum processor: 43-ghz bandwidth real-time
  amplitude measurement of 5-db squeezed light using modularized optical
  parametric amplifier with 5g technology,} {\protect\JournalTitle{Applied
  Physics Letters}} \textbf{122}, 104001 (2023).

\bibitem{Kawasaki2024}
A.~Kawasaki, R.~Ide, H.~Brunel, \emph{et~al.}, \enquote{Broadband generation
  and tomography of non-gaussian states for ultra-fast optical quantum
  processors,} {\protect\JournalTitle{Nat Commun}} \textbf{15}, 9075 (2024).

\bibitem{Kawasaki2025}
A.~Kawasaki, H.~Brunel, R.~Ide, \emph{et~al.}, \enquote{Real-time observation
  of picosecond-timescale optical quantum entanglement towards ultrafast
  quantum information processing,} {\protect\JournalTitle{Nature Photonics}} p.
  271–276 (2025).

\bibitem{Braunstein2005-2}
S.~L. Braunstein, \enquote{Squeezing as an irreducible resource,}
  {\protect\JournalTitle{Physical Review A}} \textbf{71}, 055801 (2005).

\bibitem{Zelmon1997}
D.~E. Zelmon, D.~L. Small, and D.~Jundt, \enquote{Infrared corrected sellmeier
  coefficients for congruently grown lithium niobate and 5 mol\% magnesium
  oxide –doped lithium niobate,} {\protect\JournalTitle{Journal of the
  Optical Society of America B}} \textbf{14}, 3319--3322 (1997).

\bibitem{Malitson1965}
I.~H. Malitson, \enquote{Interspecimen comparison of the refractive index of
  fused silica,} {\protect\JournalTitle{Journal of the Optical Society of
  America}} \textbf{55}, 1205--1209 (1965).

\end{thebibliography}

\end{document}